\documentclass[twocolumn]{aastex6}
 \RequirePackage{lineno}
 \usepackage{amsmath}
 \usepackage{mathrsfs}
 \usepackage{amssymb}
 \usepackage{amsthm}
 \usepackage{natbib,aasdefs,url,bm}
 \usepackage{array}
 \usepackage{float}
 \usepackage{graphicx}
 \usepackage{subfigure}
 \usepackage{color}

\newcommand{\msun}{\rm M_{\odot}}

\begin{document}

\title{High-Energy Neutrino Emission from Short Gamma-Ray Bursts:\\ 
Prospects for Coincident Detection with Gravitational Waves}

\author{Shigeo S. Kimura\altaffilmark{1,2,3}, Kohta Murase\altaffilmark{1,2,3,4}, Peter M\'{e}sz\'{a}ros\altaffilmark{1,2,3}, Kenta Kiuchi\altaffilmark{4}}
\altaffiltext{1}{Department of Physics, Pennsylvania State University, University Park, Pennsylvania 16802, USA}
\altaffiltext{2}{Department of Astronomy \& Astrophysics, Pennsylvania State University, University Park, Pennsylvania 16802, USA}
\altaffiltext{3}{Center for Particle and Gravitational Astrophysics, Pennsylvania State University, University Park, Pennsylvania 16802, USA}
\altaffiltext{4}{Center for Gravitational Physics, Yukawa Institute for Theoretical Physics, Kyoto, Kyoto 606-8502, Japan}
%
\begin{abstract}
We investigate current and future prospects for coincident detection of high-energy neutrinos and gravitational waves (GWs). 
Short gamma-ray bursts (SGRBs) are believed to originate from mergers of compact star binaries involving neutron stars. 
We estimate high-energy neutrino fluences from prompt emission, extended emission, X-ray flares, and plateau emission, and show that neutrino signals associated with the extended emission are the most promising. 
Assuming that the cosmic-ray loading factor is $\sim10$ and the Lorentz factor distribution is lognormal, we calculate the probability of neutrino detection from extended emission by current and future neutrino detectors, and find that the quasi-simultaneous detection of high-energy neutrinos, gamma rays, and GWs is possible with future instruments or even with current instruments for nearby SGRBs having extended emission. We also discuss stacking analyses that will also be useful with future experiments such as IceCube-Gen2.
\end{abstract}


\section{Introduction} \label{sec:intro}
Gravitational waves (GWs) from binary black hole (BH) mergers have been detected by the advanced LIGO (aLIGO)~\citep[e.g.,][]{LIGO16e}. The GWs from binary systems including at least one neutron star (NS), i.e. NS-BH or NS-NS binaries that have been a theoretical candidate of progenitors of short gamma-ray bursts \citep[SGRBs; e.g.,][]{1986ApJ...308L..43P,ELP89a}, can be detected in the very near future. 
The electromagnetic counterparts for such binaries have actively been discussed, including SGRBs~\citep[see][for reviews]{Nak07a,Ber14a}, off-axis emission of SGRB jets, kilonovae/macronovae in the optical/infrared band~\citep[e.g.,][]{LP98a,Met16a,Tanaka:2016sbx}, radio afterglows of merger remnants~\citep{NP11a}, and possible X-ray emission from a central engine~\citep[e.g.,][]{NKN14a,KIN15a}.

The light curves of SGRBs consist of a prompt emission, followed by several components, such as an extended emission (EE), X-ray flares, and plateau emission \citep{KI15a}, which coincide with classical afterglow emission~\citep[e.g.,][]{Nak07a,GRF09a}. 
The prompt emission is attributed to internal energy dissipation inside a relativistic jet~\citep[e.g.,][]{RM94a}, whereas the classical afterglows are caused by forward shocks propagating in the circumburst medium~\citep{MR97a}, which depend on the ambient density.
While physical mechanisms driving EEs, flares, and plateaus are not well-understood, late-time emission from a long-lasting central engine has been believed to be responsible for these emissions observed in X-rays and gamma-rays~\citep[e.g.,][]{IKZ05a,MTY11a}.
The X-ray and gamma-ray observations suggest that the total energies of these late-time emissions can be comparable to their prompt emissions \citep{CMM10a,ROM13a,KBG15a}, and that the EEs and plateau emissions may be more isotropic.  

Neutrinos serve as an important messenger to probe the physics of NS-NS and NS-BH mergers. However, the detection of MeV neutrinos from compact binary mergers is difficult even with future detectors~\citep[e.g., the detection horizon for Hyper-Kamiokande is $\sim5$~Mpc;][]{SKK11a}. High-energy neutrino searches associated with GWs have been considered and performed since the initial LIGO/VIRGO era \citep{BBB11a,IceCubeLIGO14a}, and the feasibility of joint analyses on transients with a time window of $\pm500$~s was demonstrated for GW 150914 \citep[][]{IceCubeAntares16a}. 
However, the jet formation in BH-BH mergers is disputed and special conditions seem required \citep[e.g.][]{KTT17a}. Even if the jet is launched, predictions for high-energy neutrinos depend on details of the dissipation and emission mechanisms \citep{2016ApJ...822L...9M,KS16a,MRG16a}.

We here focus on the detectability of high-energy neutrinos from SGRBs. 
GRBs have been discussed as promising sources of high-energy neutrinos~\citep[e.g.,][]{WB97a,MN06b,bec08}. 
The IceCube Collaboration has put interesting limits on the parameter spaces of GRB neutrinos~\citep[e.g.,][]{IceCube17b}. 
Their conclusions mainly come from long GRBs, and limits on prompt neutrinos from SGRBs with the current statistics are weak since their number fraction and fluences are much smaller. 
However, as we show in this work, we may still have a chance to detect neutrinos from SGRBs. 
One possibility is the occurrence of a very nearby SGRB. 
Another is high-energy neutrino emission by late-time emissions such as flares and plateaus, as suggested by \citet{MN06a}.
Since IceCube GRB searches mostly focused on prompt neutrinos so far, the neutrinos from the late-time emission are not strongly constrained \citep[see e.g.][for the analysis with a longer time window]{IceCube10a}. 

In this paper, we discuss the prospects for coincident detections of high-energy neutrinos and GWs by current and future instruments such as IceCube-Gen2 \citep{Gen214a}. 
We consider various phases of neutrino emission from SGRBs, including late-time emissions such as extended emission (EE) and plateau emission, and discuss the detectability of high-energy neutrino events, assuming that SGRBs happen within the design sensitivity range of current GW experiments (aLIGO/aVIRGO/KAGRA).

\section{High-energy Neutrinos from SGRBs}\label{sec:fluence}

High-energy neutrino emission from GRBs has been studied with detailed numerical simulations, taking into account the multi-pion production and various cooling processes~\citep[e.g.,][]{MN06b,Baerwald:2010fk}. 
Effects of multi-zone have been studied in the context of prompt emission from long GRBs, which shows highly variable light curves~\citep{Bustamante:2014oka}. 
In this work, we take the simplified approach as used in \citet{HLW12a}, which is sufficient for our purpose of comparing various phases of SGRB neutrino emission. 
We use $\varepsilon_i$ for energy of particle species $i$ in the fluid-rest frame, and $E_i$ in the observer frame. 

The photon density in a dissipation region is described by a broken power-law function: $dn_\gamma/d\varepsilon_\gamma\propto(\varepsilon_\gamma/\varepsilon_{\gamma,\rm{pk}})^{-\alpha}$ for $\varepsilon_\gamma<\varepsilon_{\gamma,\rm{pk}}$ and $dn_\gamma/d\varepsilon_\gamma\propto(\varepsilon_\gamma/\varepsilon_{\gamma,\rm{pk}})^{-\beta}$ for $\varepsilon_\gamma >\varepsilon_{\gamma,\rm{pk}}$. 
The normalization is determined by the isotropic equivalent luminosity, $L_{\gamma,\rm{iso}}=4\pi{c}\Gamma^2r_{\rm{diss}}^2{U}_\gamma$ and ${U}_\gamma=\int_{\varepsilon_{\gamma,m}}^{\varepsilon_{\gamma,M}}d\varepsilon_\gamma\varepsilon_\gamma(dn_\gamma/d\varepsilon_\gamma)$, respectively, where $\varepsilon_{\gamma,m}$  ($\varepsilon_{\gamma,M}$) is the comoving minimum (maximum) photon energy. We use $\varepsilon_{\gamma,m}=0.1$~eV and $\varepsilon_{\gamma,M}=10^6$~eV, as in \citet{MN06a}. The luminosity measured in the observed energy band, $L_{\gamma,\rm{iso}}^{*}$, depends on detectors, and $L_{\gamma, \rm{iso}}$ is several times higher than $L_{\gamma,\rm{iso}}^{*}$.

\begin{table*}[tb]
\begin{center}
\caption{Used parameters (upper table) and resultant quantities (lower table). \label{tab:models}}
\begin{tabular}{|c|cccccc|}
\hline
parameters & $\Gamma$ &$L_{\gamma,\rm iso}^{*}$ [$\rm~erg~s^{-1}$] & $\mathscr{E}_{\gamma,\rm iso}^{*}$ [erg] & $r_{\rm diss}$ [cm] & $E_{\gamma,\rm pk}$ [keV] & energy band [keV] \\
\hline
EE-mod & 30 & 3$\times10^{48}$ & $10^{51}$ & $10^{14}$ & 1 & 0.3--10 \\
EE-opt & 10 & 3$\times10^{48}$ & $10^{51}$ & 3$\times10^{13}$ & 10 & 0.3--10 \\
prompt & $10^3$ & $10^{51}$ & $10^{51}$ &  3$\times10^{13}$ & 500 & 10--$10^3$ \\
flare & 30 & $10^{48}$ &  3$\times10^{50}$ &  3$\times10^{14}$ & 0.3 & 0.3--10 \\
plateau & 30 & $10^{47}$ &  3$\times10^{50}$ &  3$\times10^{14}$ & 0.1 & 0.3--10 \\
\hline
\hline
 quantities& ${B}$ [G] &$L_{\gamma,\rm iso}$ [$\rm~erg~s^{-1}$] &$\mathscr{E}_{\gamma,\rm iso}$ [erg] &$E_{p,M}$ [EeV] &$E_{\nu,\rm\mu}$ [EeV] &$E_{\nu,\rm\pi}$ [EeV] \\
\hline
EE-mod & 2.9$\times10^{3}$ & 1.2$\times10^{49}$ & 3.8$\times10^{51}$ & 21 & 0.020 & 0.28\\
EE-opt & 5.0$\times10^{4}$ & 3.4$\times10^{49}$ & 1.1$\times10^{52}$ & 6.0 & $3.9\times10^{-4}$ & 5.4$\times10^{-3}$\\
prompt & 6.7$\times10^{3}$ & 6.1$\times10^{51}$ & 6.1$\times10^{51}$ & 60 & 0.29 & 4.0 \\
flare & 5.3$\times10^{2}$ & 3.5$\times10^{48}$ & 1.0$\times10^{51}$ & 25 & 0.11 & 1.5\\
plateau & 1.8$\times10^{2}$ &  3.8$\times10^{47}$ & 1.1$\times10^{51}$ & 13& 0.33 & 4.6\\
\hline
\end{tabular}
\end{center}
\end{table*}

For cosmic rays, we use a canonical power-law spectrum, $dN_p/dE_p\propto{E_p^{-2}}$. The total energy of non-thermal protons is normalized by $\mathscr{E}_{p,\rm{iso}}=\xi_p\mathscr{E}_{\gamma,\rm{iso}}$, where $\mathscr{E}_{\gamma,\rm{iso}}$ is the isotropic equivalent photon energy and $\xi_p=10$ is the cosmic-ray loading factor~\citep{MN06b}. Note that neutrino observations of long GRBs suggest $\xi_p\lesssim3-300$, depending on emission radii~\citep{Bustamante:2014oka,IceCube17b}. 
We use $E_{p,m}=\Gamma\varepsilon_{p,m}=\Gamma(10m_pc^2)$. 
The maximum energy is determined by the balance between the acceleration and cooling processes:
\begin{equation}
t_{\rm{acc}}^{-1}>t_{p,\rm cool}^{-1}\equiv{t_{\rm dyn}^{-1}}+t_{p,\rm syn}^{-1}+t_{p\gamma}^{-1}
\end{equation}
The acceleration time is estimated to be $t_{\rm{acc}}=\varepsilon_p/(ceB)$, where $B=\sqrt{2L_{\rm{iso}}\xi_B/(c\Gamma^2r_{\rm{diss}}^2)}$ is the comoving magnetic field strength (where $\xi_B$ is the energy fraction of the magnetic field compared to the radiation energy). For the cooling processes, we consider adiabatic cooling, synchrotron cooling, and photomeson production. The adiabatic cooling time is similar to the dynamical time: $t_{\rm{dyn}}=r_{\rm{diss}}/(c\Gamma)$. The synchrotron time for particle species $i$ is $t_{i,\rm{syn}}=6\pi{m_i^4}c^3/(m_e^2\sigma_TB^2\varepsilon_i)$, where $\sigma_T$ is the Thomson cross section. The photomeson cooling rate is evaluated by
\begin{equation}
 t_{p\gamma}^{-1}=\frac{c}{2\gamma_p^2}\int_{\overline{\varepsilon}_{\rm th}}^\infty{d}\overline{\varepsilon}_\gamma\sigma_{p\gamma}\kappa_{p\gamma}\overline{\varepsilon}_\gamma\int_{\overline\varepsilon_\gamma/(2\gamma_p)}^\infty{d}\varepsilon_\gamma \varepsilon_\gamma^{-2}\frac{dn}{d\varepsilon_\gamma},
\end{equation}
where $\gamma_p=\varepsilon_p/(m_pc^2)$, $\overline\varepsilon_{\rm{th}}\simeq145$~MeV is the threshold energy for the photomeson production, $\overline{\varepsilon}_\gamma$ is the photon energy in the proton rest frame, and $\sigma_{p\gamma}$ and $\kappa_{p\gamma}$ are the cross section and inelasticity for photomeson production, respectively. To take into account the energy dependences of $\sigma_{p\gamma}$ and $\kappa_{p\gamma}$, we use the fitting formulae based on GEANT4 \citep[see][]{MN06b}. 

\begin{figure}[tb]
\begin{center}
\includegraphics[width=\linewidth]{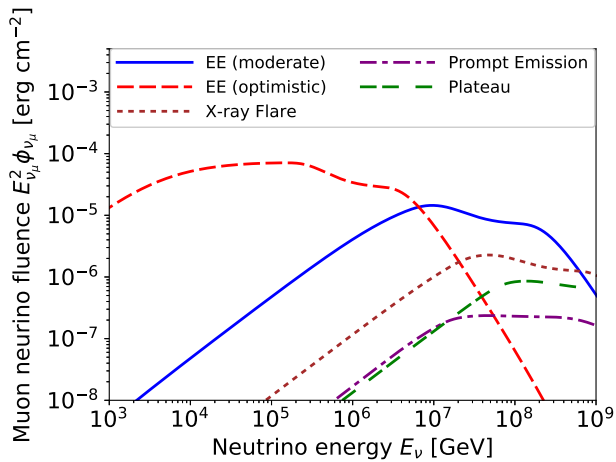}
\caption{Neutrino fluences from the EE-mod, EE-opt, prompt emission, flare, and plateau for $d_L=300$~Mpc.}
\label{fig:fluence}
\end{center}
\end{figure}

Pions generated through the photomeson production decay into muons and muon neutrinos. Using the meson production efficiency, $f_{p\gamma}\equiv{t_{p,\rm{cool}}}/t_{p\gamma}$ (which always satisfies $f_{p\gamma}<1$ in this definition~\footnote{Note that ${\rm min}[1,f_{p\gamma}]$ should be used if the photomeson production optical depth is given by $f_{p\gamma}\approx t_{\rm dyn}/t_{p\gamma}$.}), the muon neutrino spectrum produced by pion decay is estimated to be 
\begin{equation}
E_{\nu_\mu}^2\frac{dN_{\nu_\mu}}{dE_{\nu_\mu}}\approx\frac18f_{p\gamma}f_{\rm sup\pi}{E_p^2}\frac{dN_p}{dE_p},\label{eq:nu_mu}
\end{equation}
where $E_{\nu_\mu}\approx0.05E_p$ and $f_{{\rm{sup}}\pi}=1-\exp(-t_{\pi,\rm{cool}}/t_{\pi,\rm{dec}})$ is the suppression factor due to the cooling of pions. Here, $t_{\pi,\rm{dec}}=\gamma_\pi \tau_{\pi}$ is the decay time of pions ($\gamma_\pi=\varepsilon_\pi/(m_\pi{c^2})$ and $\tau_{\pi}=2.6\times10^{-8}$~s) and $t_{\pi,\rm{cool}}^{-1}=t_{\pi,\rm{syn}}^{-1}+t_{\rm{dyn}}^{-1}$ is the cooling time for pions. 
This cooling makes a spectral break in the neutrino spectrum around $E_{\nu,\pi}=\sqrt{3\pi{m_\pi^5}c^5\Gamma^2/(8m_e^2\sigma_TB^2\tau_{\pi})}$. The muons produced by the pions decay into neutrinos and positrons. The spectra of these neutrinos ($\nu_e$ and $\overline\nu_\mu$) are estimated to be
\begin{equation} 
E_{\nu_e}^2\frac{dN_{\nu_e}}{dE_{\nu_e}}\approx E_{\overline\nu_\mu}^2\frac{dN_{\overline\nu_\mu}}{dE_{\overline\nu_\mu}}\approx\frac{1}{8}f_{p\gamma}f_{\rm sup\pi}f_{\rm sup\mu}E_p^2\frac{dN_p}{dE_p}\label{eq:nu_e} 
\end{equation}
where $E_{\nu_e}\approx{E_{\overline\nu_\mu}}\approx0.05E_p$ and $f_{{\rm{sup}}\mu}$ is the suppression factor for muons.
The break for neutrino spectrum by muon cooling appears around $E_{\nu,\mu}=\sqrt{3\pi m_\mu^5c^5\Gamma^2/(8m_e^2\sigma_TB^2\tau_{\mu})}$.
The neutrino spectrum measured at the Earth is different from that at the sources due to neutrino mixing. Using the tri-bimaximal mixing matrix, the fluences are calculated via~\citep[e.g.,][]{HPS02a}
\begin{equation}
 \phi_{\nu_e+\overline\nu_e}=\frac{10}{18}\phi_{\nu_e+\overline\nu_e}^0+\frac{4}{18}(\phi_{\nu_\mu+\overline\nu_\mu}^0+\phi_{\nu_\tau+\overline\nu_\tau}^0),
\end{equation}
\begin{equation}
 \phi_{\nu_\mu+\overline\nu_\mu}=\frac{4}{18}\phi_{\nu_e+\overline\nu_e}^0+\frac{7}{18}(\phi_{\nu_\mu+\overline\nu_\mu}^0+\phi_{\nu_\tau+\overline\nu_\tau}^0),
\end{equation}
where $\phi_i^0=(dN_i/dE_i)/(4\pi{d_L^2})$ is the neutrino fluence at the source and $d_L$ is the luminosity distance.

We calculate $\phi_{\nu}$ from EEs (two cases), a prompt emission, a flare, and a plateau, whose parameters are tabulated in Table \ref{tab:models}.
The observations of SGRBs give us typical values for several parameters (see e.g. \citet{NGG11a,FBM15a,SwiftGRB16a} for prompt emissions, \citet{Swift11a,KYS15a,KBG15a,KIS17a} for EEs, \citet{CMM10a,MCG11a} for flares, and \citet{EBP09a,ROM13a,KIS17a} for plateaus), but we should note the substantial uncertainties.
The parameters that are not tabulated in the table are set to $\alpha=0.5$, $\beta=2.0$, $\xi_p=10$, $\xi_B=0.1$, and $d_L=300$~Mpc. 
This $d_L$ corresponds to the declination-averaged design sensitivity range of aLIGO for NS-NS mergers in face-on inclination~\citep{Sch11a}. 
In table \ref{tab:models}, we also tabulate the resultant physical quantities; $B$, $L_{\gamma,\rm{iso}}$, $\mathscr{E}_{\gamma,\rm{iso}}$, $E_{p,M}$, $E_{\nu,\mu}$, and $E_{\nu,\pi}$.

Figure \ref{fig:fluence} shows $\phi_{\nu_\mu}$ for the models tabulated in Table \ref{tab:models}. We see that EEs achieve much higher fluences than the others. The meson production efficiency reaches almost unity at $\sim10$~PeV ($\sim$10~TeV) for EE-mod (EE-opt), owing to their high photon number density. This makes EEs more luminous than the others.
The magnetic fields are so strong that spectral breaks due to both the muon and pion cooling supressions are seen in Figure \ref{fig:fluence}.
The proton maximum energy is determined by the photomeson production, leading to relatively lower values of $E_{p,M}$. For the other three models, $f_{p\gamma}<1$ is satisfied and the lower fluences are obtained. 
The magnetic fields are so weak that pion cooling is not important in these models. The maximum energy is determined by adiabatic losses for prompt and plateau emissions, and by photomeson production for flares.

For flares and plateaus, $\Gamma\sim10$ and $r_{\rm{diss}}\sim10^{13}$ cm are also possible~\citep[e.g.,][]{NHS14a,KIN15a}, and then, they can be as bright as EEs owing to the high pion production efficiency. 
Also, neutrino fluences from prompt emission can be higher than the plateau and flares if $\Gamma\lesssim300$ is realized.

\section{Probability of Neutrino Detection}\label{sec:detect}
\begin{table*}[tb]
\begin{center}
\caption{The detection probabilities, $P(\mathcal{N}_\mu\ge{k})$ for $d_L=300$~Mpc. IC: IceCube, Gen2: IceCube-Gen2, up+hor: upgoing + horizontal events, down: downgoing events, all: covering-factor-weighted average over the up+hor and down, $A_{\rm eff,ave}$: using the declination-averaged effective area. 
\label{tab:prob}}
\begin{tabular}{|c|ccccc|}
\hline
 EE-mod-dist-A  & IC (up+hor) & IC (down) & IC (all) & Gen2 (all) & IC ($A_{\rm eff,ave}$) \\
\hline
$P({\mathcal{N}_\mu}\ge1)$ & 0.07 & 0.04 & 0.06 & 0.21 & 0.06\\
$P({\mathcal{N}_\mu}\ge2)$ & 0.01 & 0.00 & 0.00 & 0.05 & 0.00\\
\hline
\hline
 EE-mod-dist-B  & IC (up+hor) & IC (down) & IC (all) & Gen2 (all) & IC ($A_{\rm eff,ave}$) \\
\hline
$P({\mathcal{N}_\mu}\ge1)$ & 0.11 & 0.04 & 0.08 & 0.25 & 0.08\\
$P({\mathcal{N}_\mu}\ge2)$ & 0.02 & 0.00 & 0.01 & 0.10 & 0.01\\
\hline
\hline
 EE-opt-dist-A  & IC (up+hor) & IC (down) & IC (all) & Gen2 (all) & IC ($A_{\rm eff,ave}$) \\
\hline
$P({\mathcal{N}_\mu}\ge1)$ &  0.74 & 0.25 & 0.52 & 0.86 & 0.59\\
$P({\mathcal{N}_\mu}\ge2)$ &  0.42 & 0.04 & 0.25 & 0.69 & 0.24\\
\hline
\hline
 EE-opt-dist-B  & IC (up+hor) & IC (down) & IC (all) & Gen2 (all) & IC ($A_{\rm eff,ave}$) \\
\hline
$P({\mathcal{N}_\mu}\ge1)$ & 0.60 & 0.19 & 0.41 & 0.73 & 0.47\\
$P({\mathcal{N}_\mu}\ge2)$ & 0.31 & 0.02 & 0.18 & 0.55 & 0.17\\
\hline
\end{tabular}
\end{center}
\end{table*}

The expected number of $\nu_\mu$-induced events is estimated to be
\begin{equation}
 \overline{\mathcal{N}_\mu}=\int\phi_\nu{A}_{\rm{eff}}(\delta,~E_\nu)dE_\nu,
\end{equation}
where $A_{\rm{eff}}$ is the effective area. The effective areas of upgoing+horizontal and downgoing tracks for IceCube is shown in \citet{IceCube17b} as a function of $E_\nu$.
For upgoing+horizontal muon neutrino events ($\delta>-5^\circ$), the atmospheric muons are shielded by the Earth.
For IceCube-Gen2, we use $10^{2/3}$ times larger effective areas than those of both upgoing+horizontal and downgoing events for IceCube. The effective area of downgoing muon neutrino events in IceCube-Gen2 may not be simply scaled, but the simple scaling is sufficient for the demonstrative purpose of this work. We set the threshold energy for neutrino detection to 100~GeV for IceCube and 1~TeV for IceCube-Gen2. 

The probability of detecting $k$ neutrino events, $p_k$, is described by the Poisson distribution. 
The detection probability of more than $k$ neutrinos is represented as $p(\mathcal{N}_\mu\ge{k})=1-\sum_{i<k}p_i$.
We find that for EE-mod ($\Gamma=30$), the probability for upgoing+horizontal events, $p(\mathcal{N}_\mu\ge1)$, is 0.04 and 0.16 with IceCube and IceCube-Gen2, respectively. For EE-opt ($\Gamma=10$), $ \overline{\mathcal{N}_\mu}\simeq$1.7 and 7.9 with IceCube and IceCube-Gen2, respectively.
It is possible for IceCube to detect neutrinos from EEs, while detections with IceCube-Gen2 are more promising. 
However, for $d_L=300$~Mpc, the neutrino detection for the prompt, flare, and plateau neutrino emissions may still be challenging even with IceCube-Gen2, since $p(\mathcal{N}_\mu\ge1)$ for them is less than 0.01.

The neutrino fluence of GRBs is sensitive to the Lorentz factor. To take this effect into account in a reasonable manner, we consider the distribution of $\Gamma$ to calculate the detection probability of EEs by current and future neutrino experiments.
The Lorentz factor distribution is assumed to be lognormal:
\begin{equation}
F(\Gamma)=\frac{dN_\Gamma}{d\ln\Gamma}=F_0\exp\left(-\frac{(\ln(\Gamma/\Gamma_0))^2}{2(\ln(\sigma_\Gamma))^2}\right),\label{eq:jetdist}
\end{equation}
where $F_0$ is the normalization factor ($\int_{\Gamma_{\rm min}}^{\infty}F(\Gamma)d\ln\Gamma=1$), $\Gamma_0$ is the mean Lorentz factor and $\sigma_\Gamma$ is the dispersion in logarithmic space\footnote{Although the exact shape of $F(\Gamma)$ is uncertain, the results of some analyses look lognormal, rather than Gaussian \citep{GHA04a,LYZ10a}}. 
Here, we introduce the minimum Lorentz factor $\Gamma_{\rm min}\approx2$, below which we assume that such a slow jet does not exist. We calculate $\overline{\mathcal{N}_\mu}$ for EEs with various $\Gamma$, and estimate the detection probabilities $P_k=\int{d\Gamma}F_\Gamma{p_k}$ and $P(\mathcal{N}_\mu\ge{k})=1-\sum_{i<k}P_i$.
Note that $p_k$ is a function of $\Gamma$ and $\delta$ through $\phi_\nu$ and $A_{\rm{eff}}$, respectively.
We calculate $P_k$ for upgoing+horizontal and downgoing events separately, and consider a covering-factor-weighted average as the all-sky detection probability.
Since several parameters are uncertain, we consider moderate (EE-mod-dist) and optimistic (EE-opt-dist) models. 
The basic parameters for EE-mod-dist (EE-opt-dist) are the same as those for EE-mod (EE-opt) with $\Gamma_0=30~(\Gamma_0=10)$. 
In each case, we examine $\sigma_\Gamma=2$ (EE-mod-dist-A and EE-opt-dist-A) and $\sigma_\Gamma=4$ (EE-mod-dist-B and EE-opt-dist-B).

The resultant $P_k$ are shown in Table \ref{tab:prob}, where we use $d_L=300$~Mpc.
The upgoing+horizontal events have higher probability than the downgoing events owing to a higher $A_{\rm eff}$ for low $E_\nu$. 
In EE-mod-dist cases, the lower $\sigma_\Gamma$ model (EE-mod-dist-A) has slightly lower detection probabilities, because they have a smaller fraction of lower-$\Gamma$ EEs. 
On the other hand, EE-opt-dist-A has higher detection probabilities than EE-opt-dist-B due to a smaller fraction of higher-$\Gamma$ EEs. 
We also estimate $P_k$ using declination-averaged effective area for IceCube, $A_{\rm{eff,ave}}=\int{d\Omega}A_{\rm{eff}}$, shown as IC ($A_{\rm eff,ave}$) in Table \ref{tab:prob}, which shows slightly higher $P(\mathcal{N}_\mu\ge1)$ for EE-opt-dist. 
Although the declination dependence of $A_{\rm{eff}}$ does not change our conclusion much, the declination dependent analysis is important for more quantitative evaluations.

\begin{figure}[tb]
\begin{center}
\includegraphics[width=\linewidth]{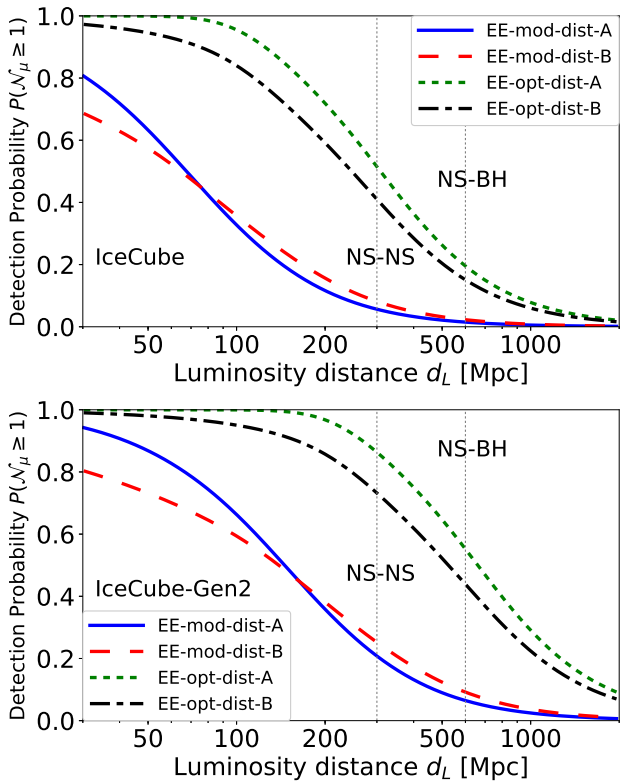}
\caption{The detection probability $P(\mathcal{N}_\mu\ge1)$ as a function of luminosity distance $d_L$. The upper and lower panels are with IceCube and IceCube-Gen2, respectively. The vertical thin-dotted lines show $d_L=300$~Mpc and $d_L=600$~Mpc.}
\label{fig:distance}
\end{center}
\end{figure}

Using the relation $\overline{\mathcal{N}_\mu}\propto\phi_{\nu_\mu}\propto{d_L^{-2}}$, 
we estimate $P(\mathcal{N}_\mu\ge1)=1-P_0$ as a function of $d_L$, which is shown in Figure \ref{fig:distance}. 
Here, we ignore the effects of cosmological redshift, since we focus on the local universe at $d_L\lesssim2$~Gpc. 
The vertical dotted lines show $d_L=300$~Mpc and $d_L=600$~Mpc, which corresponds to the sensitivity ranges of face-on NS-NS and NS-BH mergers by aLIGO, respectively.
For NS-BH mergers, since the distance is longer, $P(\mathcal{N}_\mu\ge1)$ is lower than those for NS-NS mergers.
The detection probability of nearby events is affected by $\sigma_\Gamma$, while that of distant events is not.

We estimate the detection probability within a given time interval, $\Delta{T}$, which is estimated to be $\mathcal{P}_{\Delta{T}}=1-P_0^N$, where $N$ is the number of EEs for the time interval within the covering area of neutrino detectors.
The local SGRB rate is $\sim4\rm~Gpc^{-3}~yr^{-1}-10\rm~Gpc^{-3}~yr^{-1}$ \citep[e.g.,][]{NGF06a,WP15a}, so the event rate within the sensitivity range of aLIGO (300~Mpc) is $\sim0.46\rm~yr^{-1}-1.1\rm~yr^{-1}$. 
According to the {\it Swift} results, $\sim25$~\% of SGRBs are accompaned by EEs \citep{Swift11a}, 
noting that softer instruments could detect more EEs~\citep{NKN14a}. 
Here, we simply assume that half of SGRBs have EEs, leading to $N\sim2-5$ for $\Delta{T}=10$ years. 
Within the sensitivity range of NS-BH mergers by aLIGO (600~Mpc), the SGRB rate is $\sim3.7\rm~yr^{-1}-9.0\rm~yr^{-1}$, leading to $\sim9-22$ EEs for a 5-year operation. 
The estimated values of $\mathcal{P}_{\Delta T}$ are tabulated in Table \ref{tab:delt}. 
We find that the simultaneous detection of gamma-rays, neutrinos, and GWs is possible in the era of IceCube-Gen2 and aLIGO/aVirgo/KAGRA, assuming a cosmic-ray loading factor, $\xi_p\sim10$.
This will allow us to probe the physical conditions during EEs, including the cosmic-ray loading factor and the Lorentz factor (see Section \ref{sec:summary}).

\begin{table}[tb]
\begin{center}
\caption{The detection probabilities within a given time interval, $\mathcal{P}_{\Delta T}$. The SGRB rate is assumed to be $4\rm~Gpc^{-3}~yr^{-1}-10\rm~Gpc^{-3}~yr^{-1}$
\label{tab:delt}}
\begin{tabular}{|c|cc|}
\hline
NS-NS ($\Delta T=10$ yr)  & IC (all) & Gen2 (all) \\
\hline
EE-mod-dist-A & 0.11 -- 0.25 & 0.37 -- 0.69 \\
EE-mod-dist-B & 0.16 -- 0.35 & 0.44 -- 0.77\\
EE-opt-dist-A & 0.76 -- 0.97 & 0.98 -- 1.00 \\
EE-opt-dist-B & 0.65 -- 0.93 & 0.93 -- 1.00 \\
\hline
\hline
NS-BH ($\Delta T=5$ yr)  & IC (all) & Gen2 (all) \\
\hline
EE-mod-dist-A & 0.12 -- 0.28 & 0.45 -- 0.88 \\
EE-mod-dist-B & 0.18 -- 0.39 & 0.57 -- 0.88 \\
EE-opt-dist-A & 0.85 -- 0.99 & 1.00 -- 1.00 \\
EE-opt-dist-B & 0.77 -- 0.97 & 0.99 -- 1.00 \\
\hline
\end{tabular}
\end{center}
\end{table}

In the near future, KM3NeT will be in operation. While IceCube is more suitable to observe the northern sky, KM3NeT will achieve a better sensitivity for the southern sky, helping us improve the possibility of detections.

In reality, not only $\Gamma$ but also the other parameters for EEs ($r_{\rm{diss}}$, $L_{\rm{iso}}^{\rm{obs}}$, $E_{\rm{iso}}^{\rm{obs}}$, $\alpha$, $\beta$, $E_{\gamma,\rm{pk}}$, $\xi_B$, $d_L$) should be distributed in certain ranges. 
However, their distribution functions are quite uncertain, and detailed discussion of the parameter dependences is beyond the scope of this paper.  
Systematic studies are required to obtain more solid conclusions.

\section{Summary \& Discussion}\label{sec:summary}

We have discussed the detectability of high-energy neutrinos from SGRBs that occur within the sensitivity range of GW detectors. 
We have calculated the neutrino fluences from SGRBs including prompt emission and late-time emissions (EEs, flares, plateaus), and shown that EEs may be accompanied by more efficient production of high-energy neutrinos than the other components. 
Assuming that the distribution function of the jet Lorentz factor is lognormal, the detection probability of high-energy neutrinos from EEs with IceCube and IceCube-Gen2 have been estimated as a function of $d_L$. 
Using the expected distance of GW detection from face-on NS-NS binaries ($\sim300$~Mpc), IceCube can detect neutrinos from less than 10~\% of EEs in the moderate case and around half of EEs in the optimistic case, while IceCube-Gen2 can detect around one-fourth of EEs in the moderate case and around more than three-fourth of EEs in the optimistic case, respectively. With several years of operation of IceCube-Gen2, one may expect a high probability for the quasi-simultaneous detections of gamma-rays, neutrinos, and GWs from X-ray bright SGRBs.

The sky position and timing information of a SGRB are obtained from electromagnetic waves and GWs, which allow us to reduce the atmospheric background. The intensity of the atmospheric neutrinos above TeV is around $6\times10^{-8}\rm~erg~s^{-1}~sr^{-1}~cm^{-2}$~\citep[e.g.,][]{IceCube11a}. 
Within the angular resolution of track-like events ($\sim1^\circ$) and the time window of EEs ($\sim10^2$~s), the atmospheric neutrino fluence can ideally be as small as $\sim2\times10^{-9}\rm~erg~cm^{-2}$. Although the localization accuracy can be much worse, e.g.,  $\sim5-15^\circ$ for {\it Fermi} GBM (depending on the burst duration) or a few degrees for the GW detector network (aLIGO/VIRGO/KAGRA) without electromagnetic wave counterparts~\citep[e.g.][]{Sch11a}, the atmospheric neutrino background is still much lower than the signal in many cases. Therefore, we can safely neglect the atmospheric backgrounds.

In 2030s, third-generation GW detectors such as Einstein Telescope (ET) and LIGO cosmic explorer (LIGO-CE), might be realized. ET and LIGO-CE can detect NS-NS mergers even around $z\sim2$ and $z\sim6$, respectively~\citep{ET12a,LIGO17b}.
Next-generation MeV gamma-ray satellites such as e-ASTROGAM and AMEGO are also being planned, which would be able to detect SGRBs at $z\gtrsim1$ with an angular resolution of less than a few degrees.
Since GW data can tell us a redshift of each event for given cosmological parameters~\footnote{The GW data can give the redshift and cosmological parameters independently of electromagnetic signals if the tidal effect is taken into account \citep{MR12a}.}, the redshift distribution of NS-NS mergers and SGRBs will be obtained.
In the IceCube-Gen2 era, stacking analysis are expected to be powerful. For simplicity, we assume all the EEs have the same parameters as in the EE-mod or EE-opt model, except for $d_L=5.8$ Gpc (corresponding to $z\sim0.9$). 
At this typical redshift of SGRBs~\citep{WP15a}, the SGRB rate is increased to $\sim45\rm~Gpc^{-3}~yr^{-1}$, but the atmospheric neutrinos are still negligible partially because the signal fluxes expected in this work typically have peak energies of $>10$~TeV\footnote{The temporal information of gamma-ray light curves is also useful to reduce the atmospheric background~\citep{BM14a}. See also \cite{Bustamante:2014oka}.}.
Under the assumption that half of SGRBs are accompanied by EEs, we expect $\sim1300$ EEs per year in the northern sky.
The expected number of $\nu_\mu$-induced upgoing tracks in IceCube-Gen2 is $\overline{\mathcal{N}_\mu}\simeq4.6\times10^{-4}$ and $\overline{\mathcal{N}_\mu}\simeq0.021$ for the EE-mod and EE-opt models, respectively. 
We find that the detection probability for a 3-month operation, $\mathcal{P}_{0.25\rm{yr}}$, is $\simeq0.14$ for EE-mod and $\simeq0.999$ for EE-opt. Two years of operation would be enough to increase $\mathcal{P}_{1\rm{yr}}\simeq0.69$ for EE-mod.
Detailed discussion, including the effect of cosmological evolution and parameter dependence, is left for future work.
We encourage stacking analyses specialized on not only long GRBs but also SGRBs with longer time windows in order to constrain high-energy neutrino emission associated with the late-time activities. 

High-energy neutrinos can serve as a powerful probe of cosmic-ray acceleration in SGRBs and physics of SGRB jets associated with NS-NS mergers. They can provide important clues to an outflow associated with late-time activities, whose mechanisms are highly uncertain. Several scenarios for late-time activities have been proposed to explain EEs, flares, and plateaus. 
For example, the fragmentation of the accretion disk \citep{PAZ06a} and its magnetic barrier \citep{LLG12a} may lead to a considerable amount of baryons around the central engine, which may result in a high baryon loading factor. 
On the other hand, baryon loading factors can be very low if the outflow is largely Poynting-dominated. This could be realized by not only Blandford-Znajek jets from a BH~\citep{NKN14a,KIN15a} but also a long-lived remnant NS \citep[e.g.,][]{DL98a}. Such a long-lived hypermassive NS may also emit GWs that could be detected in the future \citep[see][and references therein]{BBM13a}.

Note that all NS-NS mergers or NS-BH mergers are not necessarily accompanied by SGRBs. 
The general relativistic simulations of NS-NS mergers have revealed that a substantial amount of material ($\sim10^{-3}-10^{-2}\msun$) are ejected during the merger events~\citep{HKS13a}. If the jet energy is somewhat lower and/or the opening angle is wider, the jet cannot breakout from the merger ejecta, resulting in failed SGRBs~\citep{NHS14a}. Such choked jets have been considered as efficient high-energy neutrino sources \citep{mi13}, which may enable us to detect GWs and neutrinos simultaneously.

\acknowledgments
S.S.K thank Imre Bartos, David Burrows, Wen-Fai Fong, and Derek Fox for useful comments. This work is partially supported by Alfred P. Sloan Foundation (K.M.), NSF Grant No. PHY-1620777 (K.M.), NASA NNX13AH50G (S.S.K. and P.M.), an IGC post-doctoral fellowship program (S.S.K), JSPS KAKENHI Grant Numbers 17H06361, 16H02183, and 15K05077 (K.K), a post K computer project (project No. 9) of Japanese MEXT (K.K.), and YITP-17-96 (K.K.). We thank the YITP workshop on electromagnetic counterparts of gravitational wave sources, which initiated this project. 




\end{document}